\documentclass[a4paper,11pt]{article}

\usepackage{microtype}

\usepackage[T1]{fontenc}
\usepackage[utf8]{inputenc}

\usepackage{amsfonts}
\usepackage{amsmath}
\usepackage{amssymb}
\usepackage{mathtools}
\usepackage[binary-units=true]{siunitx}
\usepackage[boldvectors,braket]{physymb}
\usepackage{bm}

\newcommand{\mat}[1]{\bm{\mathsf{#1}}}

\usepackage{caption}
\usepackage{subcaption}
\usepackage{txfonts}

\usepackage{multicol}
\usepackage{booktabs}

\makeatletter
\newcommand\tabfill[1]{%
\dimen@\linewidth%
\advance\dimen@\@totalleftmargin%
\advance\dimen@-\dimen\@curtab%
\parbox[t]\dimen@{#1\ifhmode\strut\fi}%
}

\newcommand{\ccol}[1]{\multicolumn{1}{c}{#1}}

\usepackage{dcolumn}
\newcolumntype{d}[1]{D{.}{.}{#1}}

\usepackage[pdftex]{graphicx}
\usepackage{tikz}
\usetikzlibrary{calc,%
                decorations.markings,%
                positioning,%
                arrows,%
                chains,%
                shapes.geometric,%
                shapes,%
                shadows,%
                fit,%
                plotmarks,%
                matrix}

\usepackage{color}
\definecolor{thered}{rgb}{0.65,0.04,0.07}
\definecolor{thegreen}{rgb}{0.06,0.44,0.08}
\definecolor{theblue}{rgb}{0.02,0.2,0.68}

\usepackage[numbers,sort]{natbib}
\usepackage[unicode=true,
            bookmarks=true,
            bookmarksnumbered=false,
            bookmarksopen=false,
            breaklinks=true,
            pdfborder={0 0 1},
            backref=false,
            colorlinks=true,
            linkcolor=black,
            urlcolor=theblue,
            citecolor=theblue,
            hyperfootnotes=false]
           {hyperref}
\hypersetup{pdftitle={},
            pdfauthor={F. D. Witherden, B. C. Vermeire, P. E. Vincent}}

\begin{document}

\title{Heterogeneous Computing on Mixed Unstructured Grids with PyFR}

\author{F. D. Witherden\footnote{Corresponding author; e-mail
    freddie.witherden08@imperial.ac.uk.}, B. C. Vermeire, P. E. Vincent\\\\
  \textit{\small Department of Aeronautics, Imperial College London, SW7
    2AZ}}
\maketitle

\begin{abstract}
  PyFR is an open-source high-order accurate computational fluid
  dynamics solver for mixed unstructured grids that can target a range
  of hardware platforms from a single codebase. In this paper we
  demonstrate the ability of PyFR to perform high-order accurate
  unsteady simulations of flow on mixed unstructured grids using
  \textit{heterogeneous} multi-node hardware. Specifically, after
  benchmarking single-node performance for various platforms, PyFR
  v0.2.2 is used to undertake simulations of unsteady flow over a
  circular cylinder at Reynolds number $3\,900$ using a mixed
  unstructured grid of prismatic and tetrahedral elements on a desktop
  workstation containing an Intel Xeon E5-2697 v2 CPU, an NVIDIA Tesla
  K40c GPU, and an AMD FirePro W9100 GPU. Both the performance and
  accuracy of PyFR are assessed. PyFR v0.2.2 is freely available under
  a 3-Clause New Style BSD license (see www.pyfr.org).
\end{abstract}

\newpage

\section{Introduction}

There is an increasing desire amongst industrial practitioners of
computational fluid dynamics (CFD) to perform scale-resolving
simulations of unsteady compressible flows within the vicinity of
complex geometries. However, current generation industry-standard CFD
software---predominantly based on first- or second-order accurate
Reynolds Averaged Navier-Stokes (RANS) technology---is not well suited
to the task. Henceforth, over the past decade there has been
significant interest in the application of high-order methods for mixed
unstructured grids to such problems. Popular examples of high-order
schemes for mixed unstructured grids include the discontinuous Galerkin
(DG) method, first introduced by Reed and Hill
\cite{reed1973triangular}, and the spectral difference (SD) methods
originally proposed under the moniker `staggered-gird Chebyshev
multidomain methods' by Kopriva and Kolias in 1996
\cite{kopriva1996conservative} and later popularised by Sun et al.
\cite{sun2007high}. In 2007 Huynh \cite{huynh2007flux} proposed the
flux reconstruction (FR) approach; a unifying framework for high-order
schemes on unstructured grids that incorporates both the nodal DG
schemes of \cite{hesthaven2008nodal} and, at least for a linear flux
function, any SD scheme. In addition to offering high-order accuracy on
unstructured mixed grids, FR schemes are also compact in space, and
thus when combined with explicit time marching offer a significant
degree of element locality. This locality makes such schemes extremely
good candidates for acceleration using either the vector units of
modern CPUs or graphics processing units (GPUs)
\cite{klockner2009nodal, castonguay2011development,witherden2014pyfr}.
There exist a variety of approaches for writing accelerated codes.
These include directive based methodologies such as OpenMP 4.0 and
OpenACC, and language frameworks such as OpenCL and CUDA.
Unfortunately, at the time of writing, there exists no single approach
which is \emph{performance portable} across all major hardware
platforms. Codes desiring cross-platform portability must therefore
incorporate support for multiple approaches. Further, there is also a
growing interest from the scientific community in \emph{heterogeneous
computing} whereby multiple platforms are employed simultaneously to
solve a problem. The promise of heterogeneous computing is improved
resource utilisation on systems with a mix of hardware. Such systems
are becoming increasingly common.

PyFR is a high-order FR code for solving the Euler and compressible
Navier-Stokes equations on mixed unstructured grids
\cite{witherden2014pyfr}. Written in the Python programming language
PyFR incorporates backends for C/OpenMP, CUDA, and OpenCL. It is
therefore capable of running on conventional CPUs, as well as GPUs from
both NVIDIA and AMD, as well as heterogeneous mixtures of the
aforementioned. The objective of this paper is to demonstrate the
ability of PyFR to perform high-order accurate unsteady simulations of
flow on mixed unstructured meshes using a heterogeneous hardware
platform---demonstrating the concept of `heterogeneous computing from a
homogeneous codebase'. Specifically, we will undertake simulations of
unsteady flow over a cylinder at Reynolds number $3\,900$ using a mixed
unstructured grid of prismatic and tetrahedral elements using a desktop
workstation containing an Intel Xeon E5-2697 v2 CPU, an NVIDIA Tesla
K40c GPU, and an AMD FirePro W9100 GPU. At the time of writing these
represent the high-end workstation offerings from the three major
hardware vendors. All are designed for workstations with support for
error-correcting code (ECC) memory and double precision floating point
arithmetic.

The paper is structured as follows. In \autoref{sec:pyfr} we provide a
brief overview of the PyFR codebase. In section \autoref{sec:cyl} we
provide details of the cylinder test case. In \autoref{sec:snglnode}
single node performance is discussed. In \autoref{sec:multnode}
multi-node heterogeneous performance is discussed, and
finally in \autoref{sec:conclusions} conclusions are drawn.

\section{PyFR}
\label{sec:pyfr}

\subsection{Overview}

For a detailed overview of PyFR the reader is referred to Witherden et
al. \cite{witherden2014pyfr}. Key functionality of PyFR v0.2.2 is
summarised in \autoref{tab:pyfr-func}. We note that PyFR achieves
platform portability via use of an innovative `backend' infrastructure.

\begin{table}
  \centering
  \caption{\label{tab:pyfr-func}Key functionality of PyFR v0.2.2.}
  \begin{tabular}{rl} \toprule
    Dimensions     & 2D, 3D \\
    Elements       & Triangles, Quadrilaterals, Hexahedra,\\
                   & Tetrahedra, Prisms\\
    Spatial orders & Arbitrary\\
    Time steppers  & Euler, RK4, RK45\\
    Precisions     & Single, Double\\
    Platforms      & CPUs, NVIDIA GPUs, AMD GPUs\\
    Inter-Node Communication  & MPI\\
    Governing Systems  & Euler, Compressible Navier-Stokes\\
    \bottomrule
  \end{tabular}
\end{table}

The majority of operations within an FR time-step can be cast as matrix
multiplications of the form
\begin{equation}
  \mat{C} \leftarrow c_1\mat{A}\mat{B} + c_2\mat{C},
\end{equation}
where $c_{1,2} \in \mathbb{R}$ are scalar constants, $\mat{A}$ is a
constant operator matrix, and $\mat{B}$ and $\mat{C}$ are row-major
state matrices.  Within the taxonomy proposed by Goto et
al. \cite{goto2008anatomy} the multiplications fall into the
block-by-panel (GEBP) category.  The specific dimensions of the operator
matrices are a function of both the polynomial order $\wp$ used to
represent the solution in each element of the domain, and the element
type.  A breakdown of these dimensions for the volume-to-surface
interpolation operator matrix $\mat{M}^0$ can be found in
\autoref{tab:mat-dims}.  In PyFR platform portability of the matrix
multiply operations is achieved by deferring to the GEMM family of
subroutines provided by a Basic Linear Algebra Subprograms (BLAS)
library for the target platform.

\begin{table}
  \centering
  \caption{\label{tab:mat-dims}Dimensions of the volume-to-surface
    interpolation operator matrix $\mat{M}^0$ at orders $\wp=1,2,3,4$
    for tetrahedral, prismatic, and hexahedral element types.}
  \begin{tabular}{rllll} \toprule
    & \multicolumn{4}{c}{$\dim{\mat{M}^0}$} \\
    \cmidrule{2-5}
    Type & \ccol{$\wp=1$} & \ccol{$\wp=2$} & \ccol{$\wp=3$} & \ccol{$\wp=4$} \\
    \midrule
    Tet & $4 \times 12$ & $10 \times 24$ & $20 \times 40$ & $\hphantom{0}35 \times
    60$ \\
    Pri & $6 \times 18$ & $18 \times 39$ & $40 \times 68$ & $\hphantom{0}75 \times
    105$ \\
    Hex & $8 \times 24$ & $27 \times 54$ & $64 \times 96$ & $125 \times
    150$ \\
    \bottomrule
  \end{tabular}
\end{table}

All other operations involved in an FR time-step are
\textit{point-wise}, concerning themselves solely with data at an
individual solution/flux point. In PyFR platform portability of these
point-wise operations is achieved via use of a bespoke domain specific
language based on the Mako templating engine \cite{mako2013}. Mako
specifications of point-wise operations are converted into
backend-specific low-level code for the target platform at runtime,
which is then compiled, linked and loaded into PyFR.

\subsection{C/OpenMP backend}

The C/OpenMP backend can be used to target CPUs from a range of vendors.
The BLAS implementation employed by PyFR for the C/OpenMP backend must
be specified by the user at runtime.  Both single- and multi-threaded
libraries are supported.  When a single-threaded library is specified
PyFR will perform its own parallelisation.  Given a state matrix
$\mat{B}$ of dimension $(K,N)$ the decomposition algorithm splits
$\mat{B}$ into $N_t$ slices of dimension $(K, N/N_t)$ where $N_t$ is the
number of OpenMP threads.  Each thread then multiplies its slice through
by $\mat{A}$ to yield the corresponding slice of $\mat{C}$.  A
visualisation of this approach is shown in \autoref{fig:par-gemm}.  For
the block-by-panel multiplications encountered in FR this strategy has
been found to consistently outperform those employed by multi-threaded
BLAS libraries.

\begin{figure}
  \centering
  \begin{equation*}
    \underbrace{\left[\vphantom{\int}\;\;\hphantom{\ldots}\;\;\middle|\;\;\hphantom{\ldots}\;\;\middle|\;\;\ldots\;\;\middle|\;\;\hphantom{\ldots}\;\;\middle|\;\;\hphantom{\ldots}\;\;\right]}_{\mat{C}}
    = \underbrace{\left[\vphantom{\int}\;\;\hphantom{\ldots}\;\;\right]}_{\mat{A}} \underbrace{\left[\vphantom{\int}\;\;\hphantom{\ldots}\;\;\middle|\;\;\hphantom{\ldots}\;\;\middle|\;\;\ldots\;\;\middle|\;\;\hphantom{\ldots}\;\;\middle|\;\;\hphantom{\ldots}\;\;\right]}_{\mat{B}}
  \end{equation*}
  \caption{\label{fig:par-gemm}How matrix multiplications are
    parallelised in the C/OpenMP backend of PyFR.}
\end{figure}

\subsection{CUDA backend}

The CUDA backend can be used to target NVIDIA GPUs of compute
capability 2.0 or later. PyCUDA \cite{klockner2012pycuda} is used to
invoke the CUDA API from Python. Matrix-multiplications are performed
using the cuBLAS library which ships as part of the CUDA distribution.
The cuBLAS library is exclusively column-major. Nevertheless it is
possible to directly multiply two row-major matrices together by
utilising the fact that
\begin{equation}
  \mat{C} = \mat{A}\mat{B} \implies
  \mat{C}^T = (\mat{A}\mat{B})^T = \mat{B}^T\mat{A}^T,
\end{equation}
and observing the effect of passing a row-major matrix to a column-major
subroutine is to implicitly transpose it.

\subsection{OpenCL backend}

The OpenCL backend can be used to target  CPUs and GPUs from a range of
vendors.  The PyOpenCL package \cite{klockner2012pycuda} is used to
interface OpenCL with Python.  OpenCL natively supports runtime code
generation.  BLAS support is provided via the open source clBLAS
library, which is primarily developed and supported by AMD.  For GPU
devices clBLAS utilises auto-tuning in order to effectively target a
wide range of architectures.  Performance is heavily dependent on the
underlying OpenCL implementation.

\subsection{Distributed memory systems}

To scale out across multiple nodes PyFR utilises the Message Passing
Interface (MPI).  By construction the data exchanged between MPI ranks
is independent of the backend being used.  A natural consequence of this
is that different MPI ranks can transparently utilise different
backends---hence enabling PyFR to run on heterogenous multi-node
systems.

\section{Cylinder Test Case}
\label{sec:cyl}

\subsection{Overview}
\label{sec:cyl-overview}

Flow over a circular cylinder has been the focus of numerous previous
experimental and numerical studies. Characteristics of the flow are
known to be highly dependent on the Reynolds number $Re$, defined as
\begin{equation}
Re = \frac{u_{\infty}D}{\nu},
\end{equation}
where $u_{\infty}$ is the free-stream fluid speed, $D$ is the cylinder
diameter, and $\nu$ is the fluid kinematic viscosity. Roshko
\cite{roshko1953on} identified a stable range between $Re = 40$ and
$150$ that is characterised by the shedding of regular laminar
vortices, as well as a transitional range between $Re = 150$ and
$300$, and a turbulent range beyond $Re = 300$. These results
were subsequently confirmed by Bloor \cite{bloor1964the}, who identified
a similar set of regimes. Later, Williamson \cite{williamson1988the}
identified two modes of transition from two dimensional to three
dimensional flow. The first, known as Mode-A instability, occurs at $Re
\approx 190$ and the second, known as Mode-B instability, occurs at $Re
\approx 260$.  The turbulent range beyond $Re = 300$ can be further
sub-classified into the shear-layer transition, critical, and
supercritical regimes as discussed in the review by Williamson
\cite{williamson1996vortex}.

In the present study we consider flow over a circular cylinder at $Re =
3\,900$, and an effectively incompressible Mach number of $0.2$. This
case sits in the shear-layer transition regime identified by Williamson
\cite{williamson1996vortex}, and contains several complex flow
features, including separated shear layers, turbulent transition, and a
fully turbulent wake.

\subsection{Domain}

In the present study we use a computational domain with dimensions
$[-9D,25D]$; $[-9D,9D]$; and $[0,\pi D]$ in the stream-, cross-, and
span-wise directions, respectively. The cylinder is centred at
$(0,0,0)$. The span-wise extent was chosen based on the results of
Norberg \cite{norberg1988ldv}, who found no significant influence on
statistical data when the span-wise dimension was doubled from $\pi D$
to $2\pi D$. Indeed, a span of $\pi D$ has been used in the majority of
previous numerical studies \cite{ma2000dynamics, breuer1998large,
kravchenko2000numerical}, including the recent DNS study of Lehmkuhl et
al. \cite{lehmkuhlthe2013}. The stream-wise and cross-wise dimensions
are comparable to the experimental and numerical values used by
Parnaudeau et al. \cite{parnaudeau2008experimental}, whose results will
be directly compared with ours. The overall domain dimensions are also
comparable to those used for DNS studies by Lehmkuhl et al.
\cite{lehmkuhlthe2013}. The domain is periodic in the span-wise
direction, with a no-slip isothermal wall boundary condition applied at
the surface of the cylinder, and Riemann invariant boundary conditions
applied at the far-field.

\subsection{Mesh}
\label{sec:cyl-meshes}

The domain was meshed in two ways. The first mesh consisted
of entirely hexahedral elements, and the second a mixture of prismatic
elements in the near wall boundary layer region, and tetrahedral
elements in the wake and far-field. Both meshes employed quadratically curved elements, and were designed to fully
resolve the near wall boundary layer region when $\wp =
4$. Specifically, the maximum skin friction coefficient was estimated
\emph{a priori} as $C_f \approx 0.075$ based on the LES results of
Breuer \cite{breuer1998large}. The height of the first element was then
specified such that when $\wp = 4$ the first solution point from
the wall sits at $y+ \approx 1$, where non-dimensional wall units are
calculated in the usual fashion as $y+ = u_{\tau} y / \nu$ with
$u_{\tau} = \sqrt{C_f/2}u_{\infty}$.

\begin{figure}
  \centering
  \begin{subfigure}[b]{.49\linewidth}
    \centering
    \includegraphics[height=3cm]{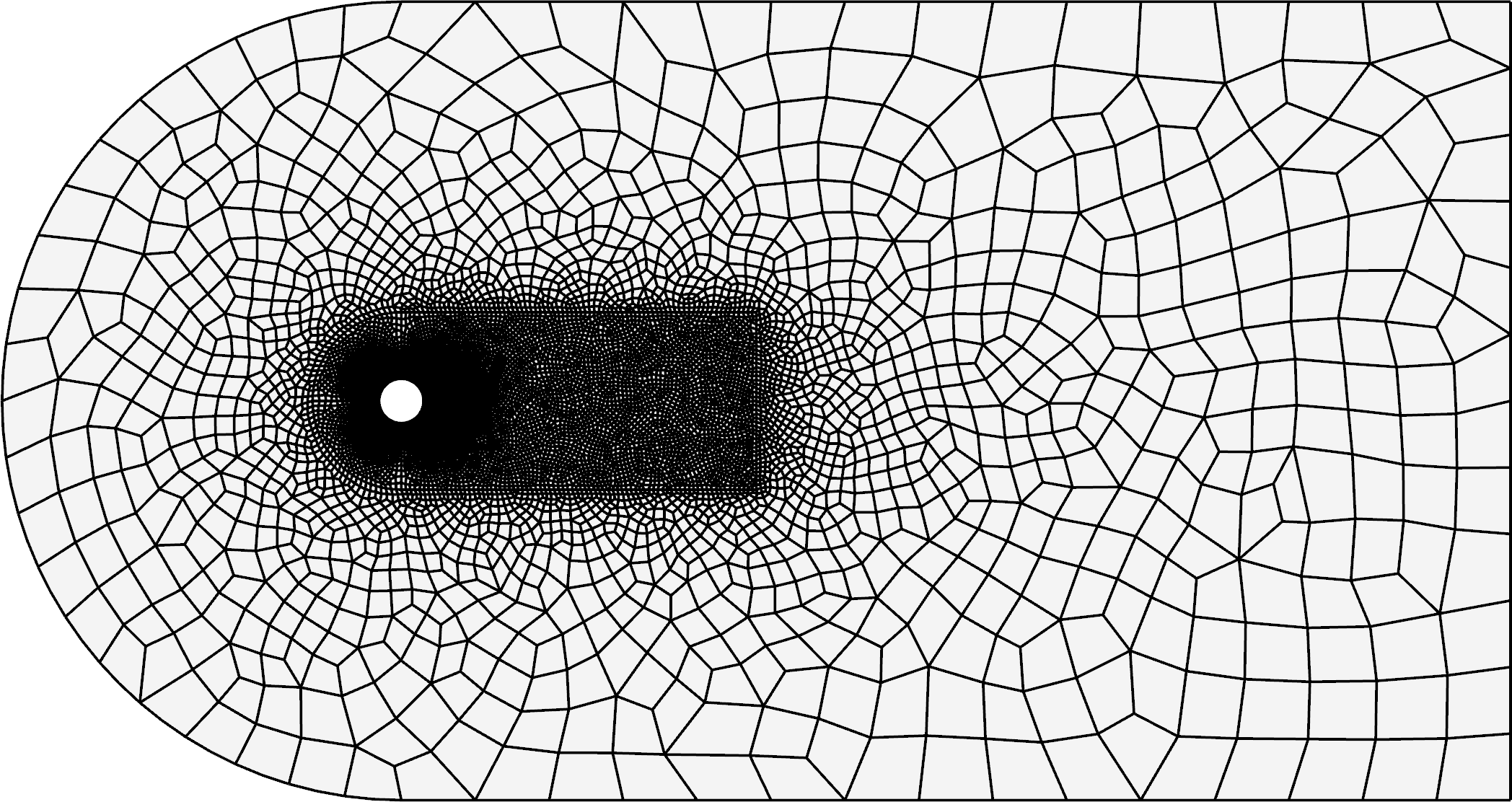}
    \caption{Hexahedral, far-field.}
  \end{subfigure}
  \begin{subfigure}[b]{.49\linewidth}
    \centering
    \includegraphics[height=3cm]{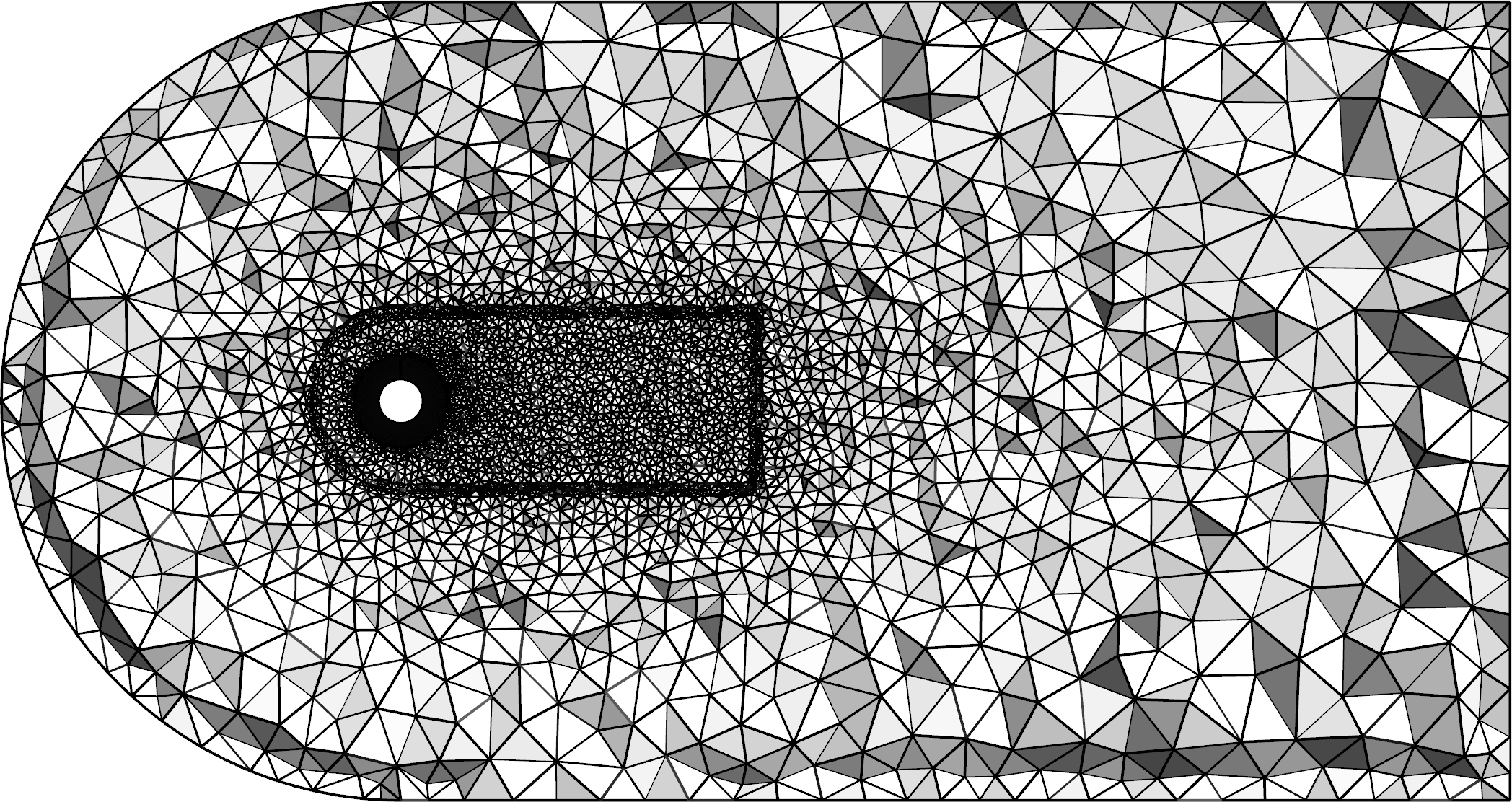}
    \caption{Prism/tetrahedral, far-field.}
  \end{subfigure}\vskip12pt
  \begin{subfigure}[b]{.49\linewidth}
    \centering
    \includegraphics[height=2.5cm]{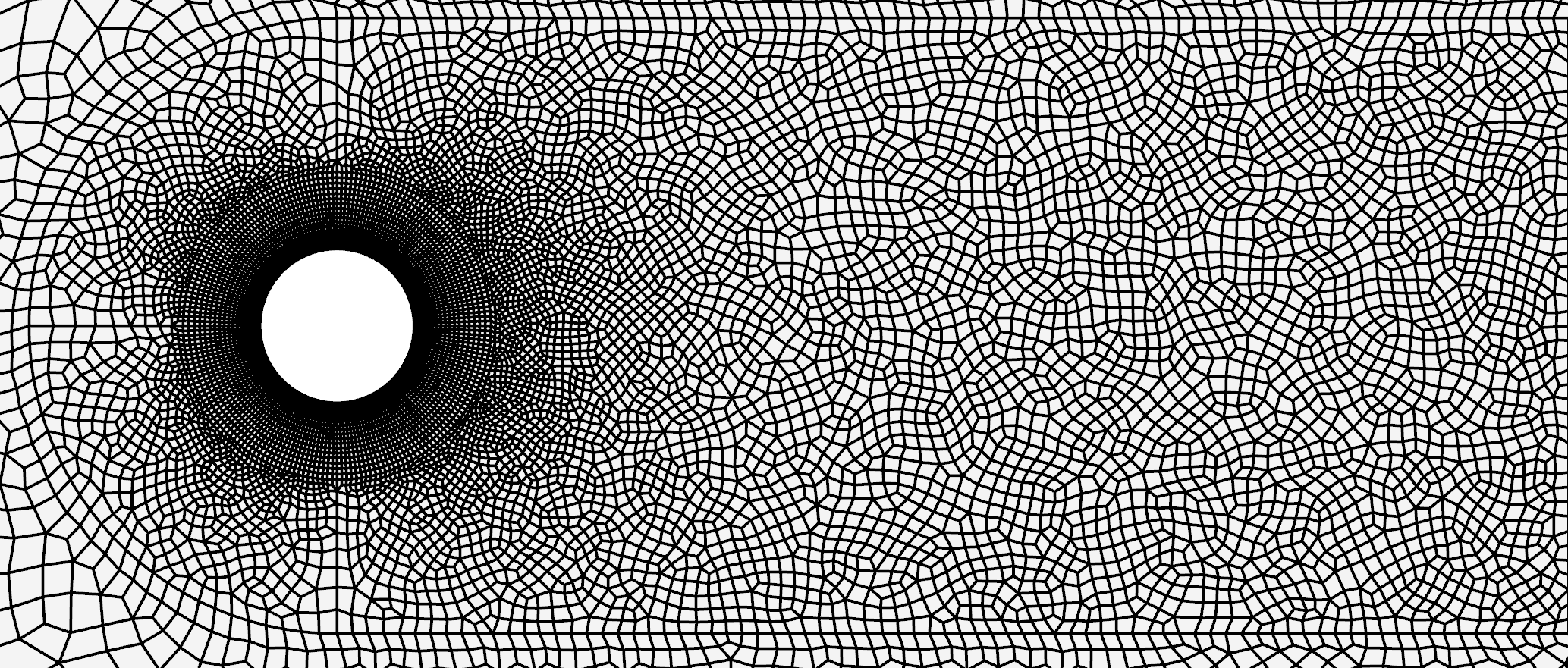}
    \caption{Hexahedral, wake.}
  \end{subfigure}
  \begin{subfigure}[b]{.49\linewidth}
    \centering
    \includegraphics[height=2.5cm]{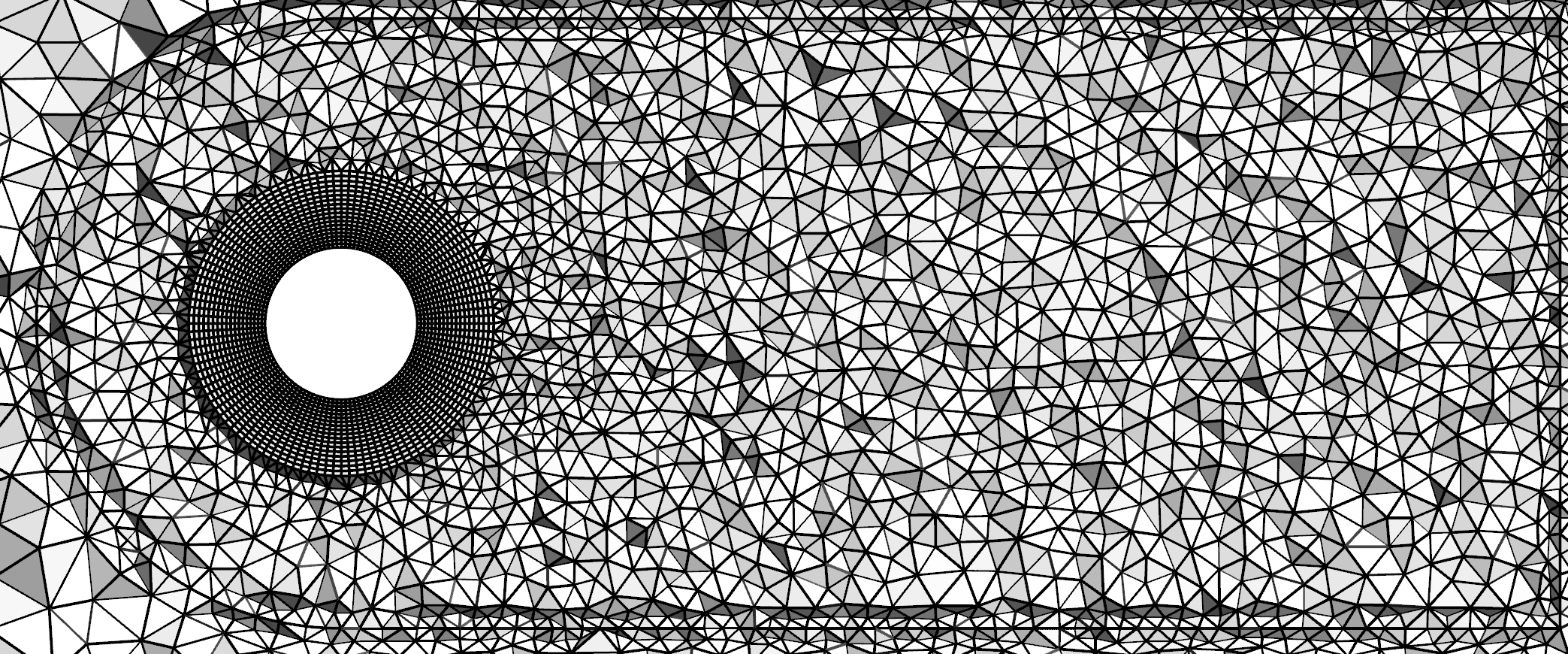}
    \caption{Prism/tetrahedral, wake.}
  \end{subfigure}
  \caption{\label{fig:meshes}Cutaways through the two meshes.}
\end{figure}

The hexahedral mesh had 200 elements in the circumferential direction,
and 10 elements in the span-wise direction, which when $\wp = 4$
achieves span-wise resolution comparable to that used in previous
studies; as discussed by Breuer \cite{breuer1998large} and the
references contained therein. The prism/tetrahedral mesh has 120
elements in the circumferential direction, and 20 elements in the
span-wise direction, these numbers being chosen to help reduce face
aspect ratios at the edges of the prismatic layer; which facilitates
transition to the fully unstructured tetrahedral elements in the
far-field. In total the hexahedral mesh contained $118\,070$ elements,
and the prism/tetrahedral mesh contained $79\,344$ prismatic elements
and $227\,298$ tetrahedral elements. Both meshes are shown in
\autoref{fig:meshes}.

\subsection{Methodology}

\begin{figure}
  \centering
  \includegraphics{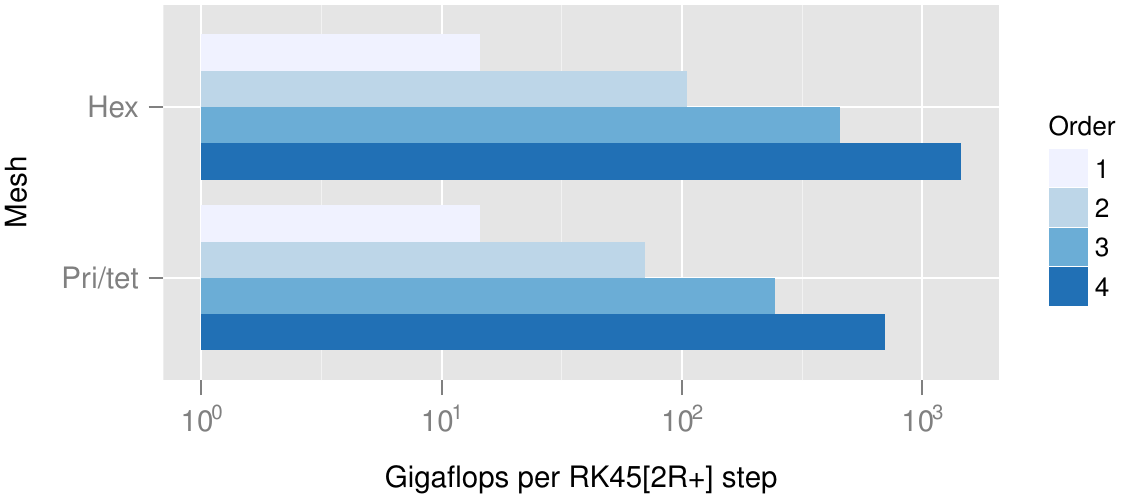}%
  \caption{\label{fig:gflops-per-step}Computational effort required for
    the $118\,070$ element hexahedral mesh and the mixed mesh with
    $79\,344$ prismatic elements and $227\,298$ tetrahedral elements.}
\end{figure}

The compressible Navier-Stokes equations, with constant viscosity, were
solved on each of the two meshes shown in \autoref{fig:meshes}. A DG
scheme was used for the spatial discretisation, a Rusanov Riemann
solver was used to calculate the inviscid fluxes at element interfaces,
and the explicit RK45[2R+] scheme of Carpenter and Kennedy
\cite{kennedy2000low} was used to advance the solution in time. No
sub-grid model was employed, hence the approach should be considered
ILES/DNS, as opposed to classical LES. Values of $\wp$ from one to four
were used with each mesh. The approximate memory requirements of PyFR
for these simulations can are detailed in \autoref{tab:mem}. The total
number of floating point operations required by an RK45[2R+] time-step
for each simulation are detailed in \autoref{fig:gflops-per-step}. When
running with $\wp = 1$ both meshes require ${\sim}1.5 \times 10^{10}$
floating point operations per time-step. This number can be seen to
increase rapidly with the polynomial order.

\begin{table}
  \centering
  \caption{\label{tab:mem}Approximate memory requirements of PyFR for
    the two cylinder meshes.}
  \begin{tabular}{rrrrr} \toprule
    & \multicolumn{4}{c}{Device memory / GiB} \\
    \cmidrule{2-5}
    Mesh & \ccol{$\wp=1$} & \ccol{$\wp=2$} & \ccol{$\wp=3$} &
    \ccol{$\wp=4$} \\
    \midrule
    Hex & $0.8$ & $2.0$ & $4.1$ & $7.2$ \\
    Pri/tet & $1.1$ & $2.6$ & $4.7$ & $7.7$ \\
    \bottomrule
  \end{tabular}
\end{table}

\section{Single-Node Performance}
\label{sec:snglnode}

\subsection{Overview}

In this section we will analyse performance of PyFR on an Intel Xeon
E5-2697 v2 CPU, an NVIDIA Tesla K40c GPU, and an AMD FirePro W9100 GPU.
These are the individual platforms used to construct the multi-node
heterogeneous system employed in \autoref{sec:multnode}.

\subsection{Hardware specifications}

Various attributes of the E5-2697, K40c, and W9100 are detailed in
\autoref{tab:plat-stats}. The theoretical peaks for double precision
arithmetic and memory bandwidth were obtained from vendor
specifications. However, in practice it is usually only possible to
obtain these theoretical peak values using specially crafted code
sequences.  Such sequences are almost always platform specific and
seldom perform useful computations.  Consequently, we also calculate and
tabulate \emph{reference} peaks. Reference peaks for double precision
arithmetic are defined here as the maximum number of giga floating point
operations per second (GFLOP/s) obtained while multiplying two large
double precision row-major matrices using DGEMM from an appropriate BLAS
library.  Reference peaks for memory bandwidth are defined here as the
rate, in gigabytes per second (GB/s), that data can be copied between
two one gigabyte buffers.  Reference peaks for the E5-2697 were obtained
using DGEMM from the Intel Math Kernel Libraries (MKL) version 11.1.2,
and with the E5-2697 paired with four DDR3-1600 DIMMs (with Turbo Boost
enabled). Reference peaks for the K40c were obtained using DGEMM from
cuBLAS as shipped with CUDA 6, with GPU boost disabled and ECC
enabled. Reference peaks for the W9100 were obtained using DGEMM from
clBLAS v2.0 with version 1411.4 of the AMD APP OpenCL runtime.  ECC
settings for the W9100 were left unchanged.

We note that on the K40c ECC is implemented in software, and when
enabled error-correction data is stored in global memory.  A consequence
of this is that when ECC is enabled there is a reduction in available
memory and memory bandwidth.  This partially accounts for the
discrepancy observed between the theoretical and reference memory
bandwidths for the K40c. We also note that for the K40c and the E5-2697,
reference peaks for double precision arithmetic are in excess of 80\% of
their theoretical values. However, for the W9100 the reference peak for
double precision arithmetic is only 34\% of its theoretical value.  This
value is not significantly improved via the auto-tuning utility that
ships with clBLAS.  It is hoped that this figure will improve with
future releases of clBLAS.

\begin{table}
  \centering
  \caption{\label{tab:plat-stats}Baseline attributes of the three
    hardware platforms. For the NVIDIA Tesla K40c GPU Boost was left
    disabled and ECC was enabled. The Intel Xeon E5-2697 v2 was paired
    with four DDR3-1600 DIMMs with Turbo Boost enabled.}
  \begin{tabular}{rrrr} \toprule
   & \multicolumn{3}{c}{Platform} \\  \cmidrule{2-4}
   & \ccol{K40c} & \ccol{W9100} & \ccol{E5-2697} \\ \midrule
   Arithmetic / GFLOP/s\hspace{1ex} \\
   theoretical peak & 1430 & 2620 & 280 \\
   reference peak & 1192 & 890 & 231 \\ \midrule
   Memory Bandwidth / GB/s\hspace{1ex} \\
   theoretical peak & 288 & 320 & 51.2 \\
   reference peak & 190 & 261 & 37.1 \\ \midrule
   Thermal Design Power / W & 235 & 275 & 130 \\
   Memory / GB & 12 & 16 &  \\
   Clock / MHz & 745 & 930 & 3000 \\
   Transistors / Billion & 7.1 & 6.2 & 4.3 \\ \bottomrule
  \end{tabular}
\end{table}

Finally, as an aside, we note that the number of `cores' available on
each platform have been deliberately omitted from
\autoref{tab:plat-stats}.  It is our contention that the term is both
ill-defined and routinely subject to abuse in the literature.  For
example, the E5-2697 is presented by Intel as having 12 cores, whereas
the K40c is described by NVIDIA as having 2880 `CUDA cores'.  However,
whereas the cores in the E5-2697 can be considered linearly independent
those in the K40c can not.  The rough equivalent of a CPU core in NVIDIA
parlance is a `streaming multiprocessor', or SMX, of which the K40c has
15.  Additionally, the E5-2697 has support for two-way simultaneous
multithreading---referred to by Intel as Hyper-Threading---permitting
two threads to execute on each core.  At any one instant it is therefore
possible to have up to 24 independent threads resident on a single
E5-2697.  The AMD equivalent of a CUDA core is a `stream processor' of
which the W9100 has 2816.  This is not to be confused with the
aforementioned streaming multiprocessor of NVIDIA; for which the AMD
equivalent is a `Compute Unit'.  Practically, both CUDA cores and stream
processors are closer to the individual vector lanes of a traditional
CPU core.  Given this minefield of confusing nomenclature we have
instead opted to simply state the peak floating point capabilities of
the hardware.

\subsection{Performance}

By measuring the wall clock time required for PyFR to take a defined
number of time-steps, and utilising the operation counts per time-step
detailed in \autoref{fig:gflops-per-step}, one can calculated the
sustained performance of PyFR in GFLOP/s when running with the meshes
detailed in \autoref{sec:cyl-meshes} and $\wp = 1,2,3,4$.

\begin{figure}
  \centering
  \includegraphics{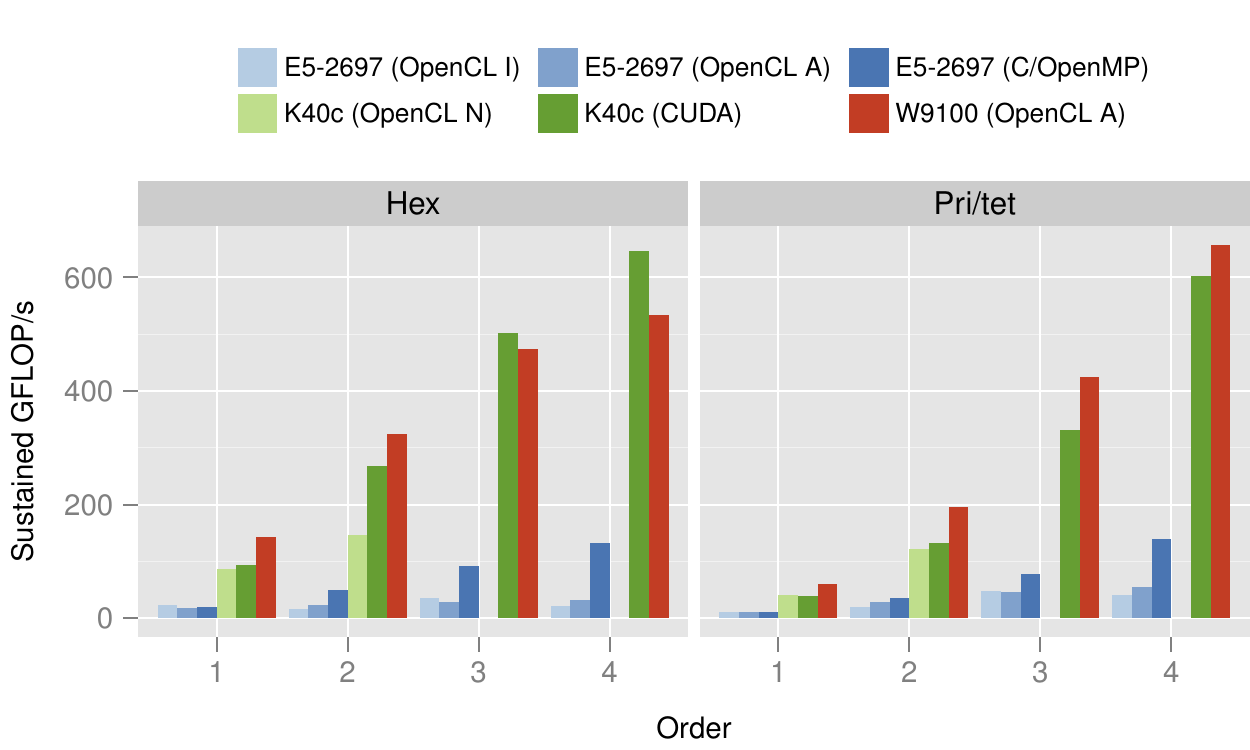}%
  \caption{\label{fig:gflops-sim}Sustained performance of PyFR in
    GFLOP/s for the various pieces of hardware.  The backend used by
    PyFR is given in parentheses.  For the OpenCL backend the initial of the
    vendor is suffixed.  As the NVIDIA OpenCL platform is limited to
    $4\,\text{GiB}$ of memory no results are available for $\wp = 3,4$.}
\end{figure}

Sustained performance of PyFR for the various hardware platforms is
shown in \autoref{fig:gflops-sim}. From the figure it is clear that the
computational efficiency of PyFR increases with the polynomial order.
This is consistent with higher order simulations having an increased
compute intensity per degree of freedom. This additional intensity
results in larger operator matrices that are better suited to the
tiling schemes employed by BLAS libraries. The OpenCL implementation
shipped by NVIDIA as part of CUDA only supports the use of 32-bit
memory pointers. As such a single context is limited to $4\,\text{GiB}$
of memory, \emph{cf.} \autoref{tab:mem}. It was therefore not possible
to perform the third and fourth order simulations for either of the two
meshes using the OpenCL backend with the K40c.

The Intel and AMD implementations of OpenCL, when used in conjunction
with clBLAS, are only competitive with the C/OpenMP backend when $\wp =
1$ for the hexahedral mesh, and $\wp = 1,2$ for the prism/tetrahedral
mesh.  This is also the case when comparing performance between the CUDA
backend and the NVIDIA OpenCL backend on the K40c.  Prior analysis by
Witherden et al. \cite{witherden2014pyfr} suggests that at these orders
a reasonable proportion of the wall clock time will be spent in the
bandwidth-bound point-wise kernels as opposed to DGEMM.  On account of
being bandwidth-bound such kernels do not extensively test the
optimisation capabilities of the compiler.  By the time $\wp = 4$ both
implementations of OpenCL on the E5-2697 are delivering between one
third and one quarter of the performance of the native backends.  This
highlights the difficulties associated with writing a performant DGEMM
routine and hence severely impacts the performance portability of the
OpenCL backend.  This result lends credence to our opening assertion
that there are currently no programming methodologies that are
performance portable across a range of platforms.  Further, it also
justifies the approach that has been adopted by PyFR.

Performance of the K40c culminates at $647\,\text{GFLOP/s}$ for the $\wp
= 4$ hexahedral mesh.  This represents some $45\%$ of the theoretical
peak and $54\%$ of the reference peak.  By comparison the E5-2697
obtains $132\,\text{GFLOP/s}$ for the same simulation equating to $47\%$
and $57\%$ of the theoretical and reference peaks, respectively.
Performance does improve slightly to $140\,\text{GFLOP/s}$ for the $\wp
= 4$ prism/tetrahedral mesh, however.  On this same mesh at $\wp = 4$
the W9100 can be seen to sustain $657\,\text{GFLOP/s}$ of throughput.
Although, in absolute terms, this observation represents the highest
sustained rate of throughput it corresponds to just $25\%$ of the
theoretical peak.  However, working in terms of realisable peaks, we
find PyFR obtaining some $74\%$ of the reference value.

\section{Multi-Node Heterogeneous Performance and Accuracy}
\label{sec:multnode}

\subsection{Overview}

Having determined the performance characteristics of PyFR on various
individual platforms, we will now investigate the ability of PyFR to
undertake simulations on a multi-node heterogeneous system containing an
Intel Xeon E5-2697 v2 CPU, an NVIDIA Tesla K40c GPU, and an AMD FirePro
W9100 GPU.

\subsection{Mesh partitioning}

In order to distribute a simulation across the nodes of the
heterogeneous system it is first necessary to partition the mesh. High
quality partitions can be readily obtained using a graph partitioning
package such as METIS \cite{karypis1998fast} or SCOTCH
\cite{pellegrini1996scotch}.

When partitioning a mixed element mesh for a \emph{homogeneous} cluster
it is necessary to suitably weight each element type according to its
computational cost.  This cost depends both upon the platform on which
PyFR is running and the order at which the simulation is being
performed.  In principle it is possible to measure this cost; however in
practice the following set of weights have been found to give
satisfactory results across most polynomial orders and platforms
\[
\text{hex} : \text{pri} : \text{tet} = 3:2:1,
\]
where larger numbers indicate a greater computational cost.  One
subtlety that arises here, is that from a graph partitioning standpoint
there is no penalty associated with placing a sole vertex (element) of a
given weight inside of a partition.  Computationally, however, there is
a very real penalty incurred from having just a single element of a
certain type inside of the partition.  It is therefore desirable to
avoid mesh partitions where any one partition contains less than around
a thousand elements of a given type.  An exception is when a partition
contains no elements of such a type---in which case zero overheads are
incurred.

When partitioning a mesh with one type of element for a
\emph{heterogeneous} cluster it is necessary to weight the partition
sizes in line with the performance characteristics of the hardware on
each node.  However, in the case of a mixed element mesh on a
heterogeneous cluster the weight of an element is no longer static but
rather depends on the partition that it is placed in---a significantly
richer problem.  Solving such a problem is currently beyond the
capabilities of most graph partitioning packages.  Accordingly, mixed
element meshes that are partitioned for heterogeneous clusters often
exhibit inferior load balancing than those partitioned for homogeneous
systems.  Moreover, for consistent performance it is necessary to
dedicate a CPU core to each accelerator in the system.  The amount of
useful computation that can be performed by the host CPU is therefore
reduced in accordance with this.

Given the single-node performance numbers of \autoref{fig:gflops-sim} it
comports to pair the E5-2697 with the C/OpenMP backend, the K40c with
the CUDA backend, and the W9100 with the OpenCL backend, in order to
achieve optimal performance.  Employing the results of
\autoref{fig:gflops-sim}, in conjunction with some light
experimentation, a set of partitioning weights were obtained and are
tabulated in \autoref{tab:part-wgts}.

\begin{table}
  \centering
  \caption{\label{tab:part-wgts}Partition weights for the multi-node
    heterogeneous simulation.}
  \begin{tabular}{rllll} \toprule
     & \multicolumn{4}{c}{$\text{E5-2697} : \text{W9100} : \text{K40c}$} \\
    \cmidrule{2-5}
    Mesh & \ccol{$\wp=1$} & \ccol{$\wp=2$} & \ccol{$\wp=3$} & \ccol{$\wp=4$} \\
    \midrule
    Hex & $3{:}27{:}23$ & $3{:}27{:}24$ & $4{:}24{:}26$ & $4{:}24{:}28$ \\
    Pri/tet & $5{:}33{:}17$ & $5{:}33{:}17$ & $5{:}30{:}20$ &
    $5{:}27{:}23$ \\
    \bottomrule
  \end{tabular}
\end{table}

\subsection{Performance}

\begin{figure}
  \centering
  \includegraphics{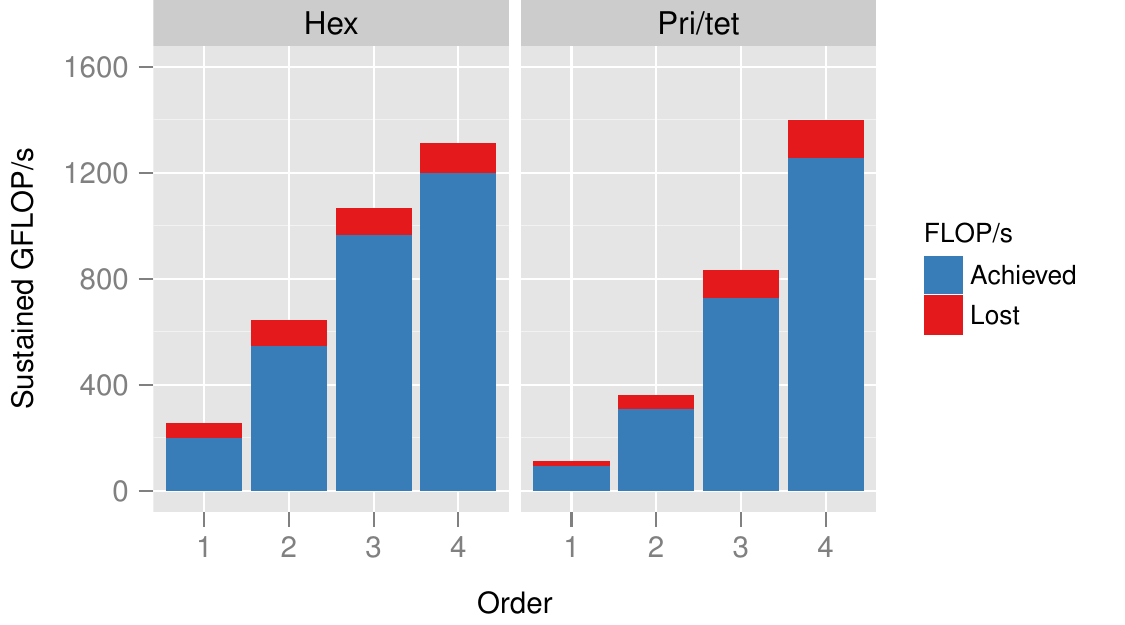}%
  \caption{\label{fig:gflops-hyb}Sustained performance of PyFR on the
    multi-node heterogeneous system for each mesh with $\wp = 1,2,3,4$.
    Lost FLOP/s represent the difference between the achieved FLOP/s and
    the sum of the E5-2697 (C/OpenMP), K40c (CUDA), and W9100 (OpenCL A)
    bars in \autoref{fig:gflops-sim}.}
\end{figure}

Sustained performance of PyFR on the multi-node heterogeneous system
for each of the meshes detailed in \autoref{sec:cyl-meshes} with
$\wp = 1,2,3,4$ is shown in \autoref{fig:gflops-hyb}. Under the
assumptions of perfect partitioning and scaling one would expect the
sustained performance of the heterogeneous simulation to be equivalent
to the sum of the E5-2697 (C/OpenMP), K40c (CUDA), and W9100 (OpenCL A)
bars in \autoref{fig:gflops-sim}. However, for reasons outlined in the
preceding paragraphs these assumptions are unlikely to hold. Some of
the available FLOP/s can therefore be considered as `lost'. For the
hexahedral mesh the fraction of lost FLOP/s varies from $22.5\%$ when
$\wp = 1$ to $8.7\%$ in the case of $\wp = 4$. With the exception of
$\wp = 1$ the fraction of lost FLOP/s are a few percent higher for the
mixed mesh. This is understandable given the additional complexities
associated with mixed mesh partitioning and can likely be improved upon
by switching to order-dependent element weighting factors.

\subsection{Accuracy}

In this section we present instantaneous and time-span-averaged
(henceforth referred to as averaged) results obtained using the
multi-node heterogeneous system and the mixed unstructured
prism/tetrahedral mesh with $\wp = 1,2,3,4$. Following the approach of
Breuer \cite{breuer1998large} all averaged results were obtained over
$100D/u_\infty$ in time, and the full span of the domain. The $\wp = 1$
simulation was initialised with spatially constant free-stream initial
conditions, and run for a lead in time of ${\sim}50D/u_\infty$ before
time-averaging began. Subsequent simulations, at $\wp >1$, were
initialised with the final solution state from the previous simulation
at $\wp-1$, and time-averaging began immediately. For all $\wp$ this
approach led to averaged results exhibiting U-shape or Mode-L
characteristics (following the terminology of Ma et al.
\cite{ma2000dynamics} and Lehmkuhl et al. \cite{lehmkuhlthe2013}
respectively). Hence, all averaged results were compared with the
experimental results of Norberg et al. \cite{norberg1988ldv} and
Parnaudeau et al. \cite{parnaudeau1998experimental}, which also
exhibited U-shape/Mode-L characteristics, as well as Mode-L results
from the recent DNS study of Lehmkuhl et al. \cite{lehmkuhlthe2013}.

\begin{figure}
  \centering
  \begin{subfigure}[b]{1\linewidth}
    \centering
    \includegraphics[width=0.6\linewidth]{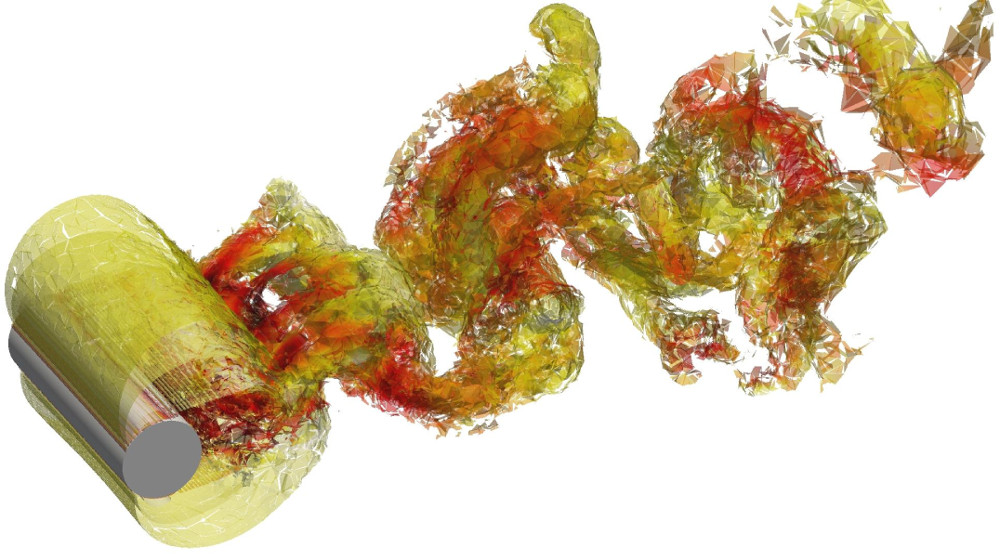}
    \caption{$\wp = 1$.}
  \end{subfigure}
  \begin{subfigure}[b]{1\linewidth}
    \centering
    \includegraphics[width=0.6\linewidth]{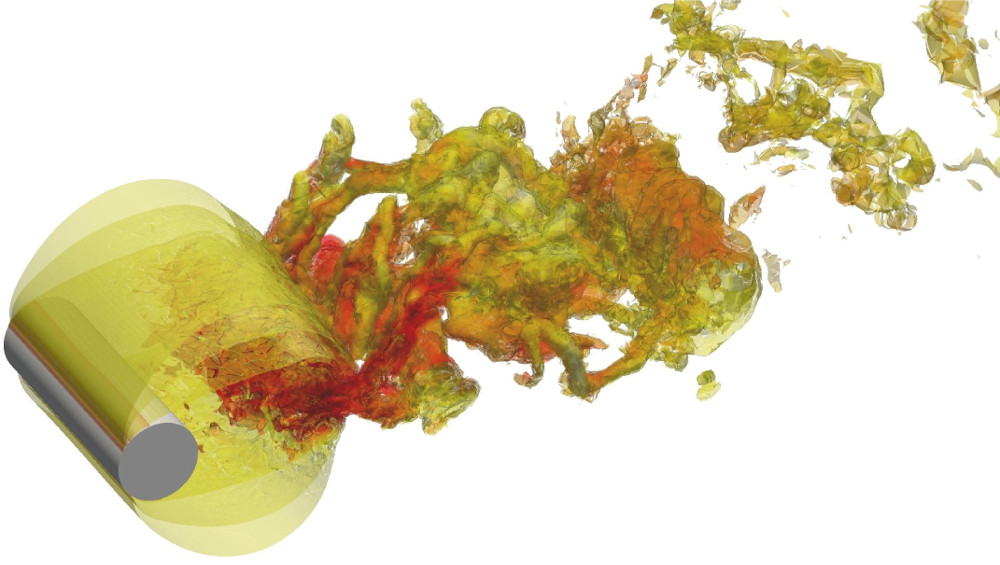}
    \caption{$\wp = 2$.}
  \end{subfigure}
  \begin{subfigure}[b]{1\linewidth}
    \centering
    \includegraphics[width=0.6\linewidth]{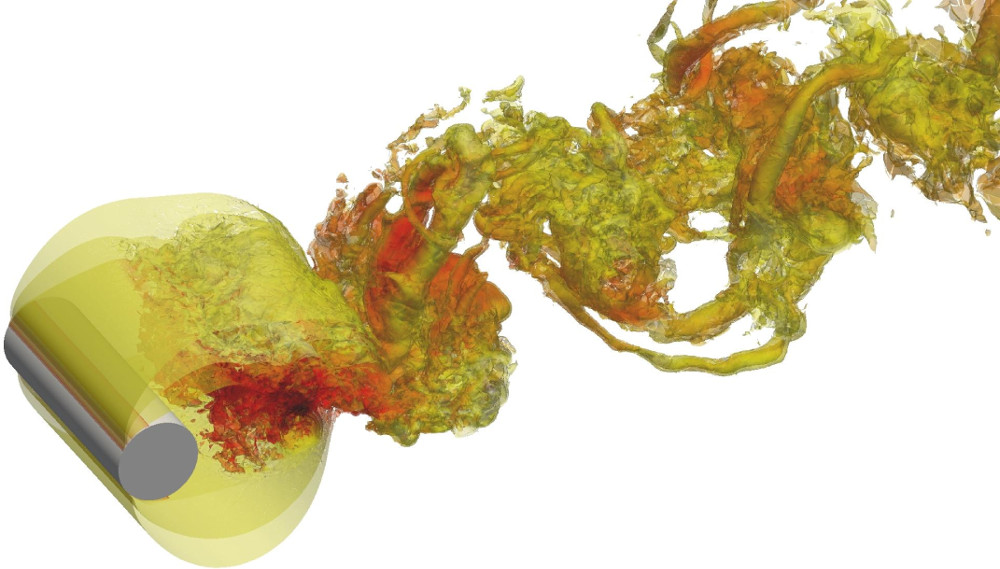}
    \caption{$\wp = 3$.}
  \end{subfigure}
  \begin{subfigure}[b]{1\linewidth}
    \centering
    \includegraphics[width=0.6\linewidth]{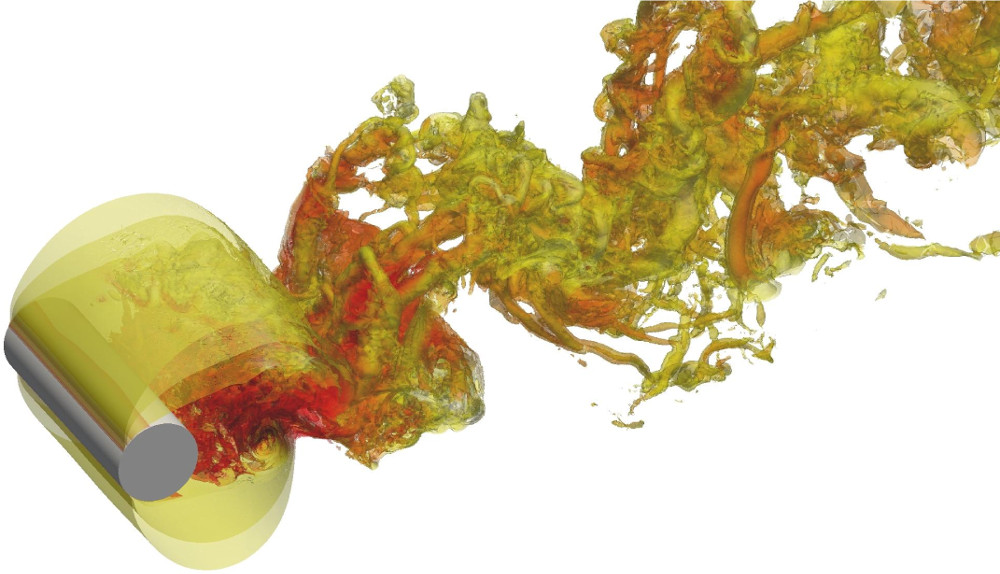}
    \caption{$\wp = 4$.}
  \end{subfigure}
  \caption{\label{fig:isosurfaces}Instantaneous surfaces of iso-density
    coloured by velocity magnitude.}
\end{figure}

Instantaneous surfaces of iso-density are shown in
\autoref{fig:isosurfaces} for each polynomial order. We observe that
the $\wp = 1$ simulation captures predominantly large-scale structures
in the turbulent wake behind the cylinder. As $\wp$ is increased we are
able to capture a greater number of small scale turbulent structures.
This is due to an increase in the total number of degrees of freedom in
the domain, as well as a relative decrease in discretisation errors for
the higher-order schemes. We observe laminar flow at the leading edge
of the cylinder for all test cases, transition near the separation
points, and finally turbulent flow in the wake region. These are the
characteristic features of the shear-layer transition regime as
described by Williamson \cite{williamson1996vortex}.

\begin{figure}
  \centering
  \includegraphics{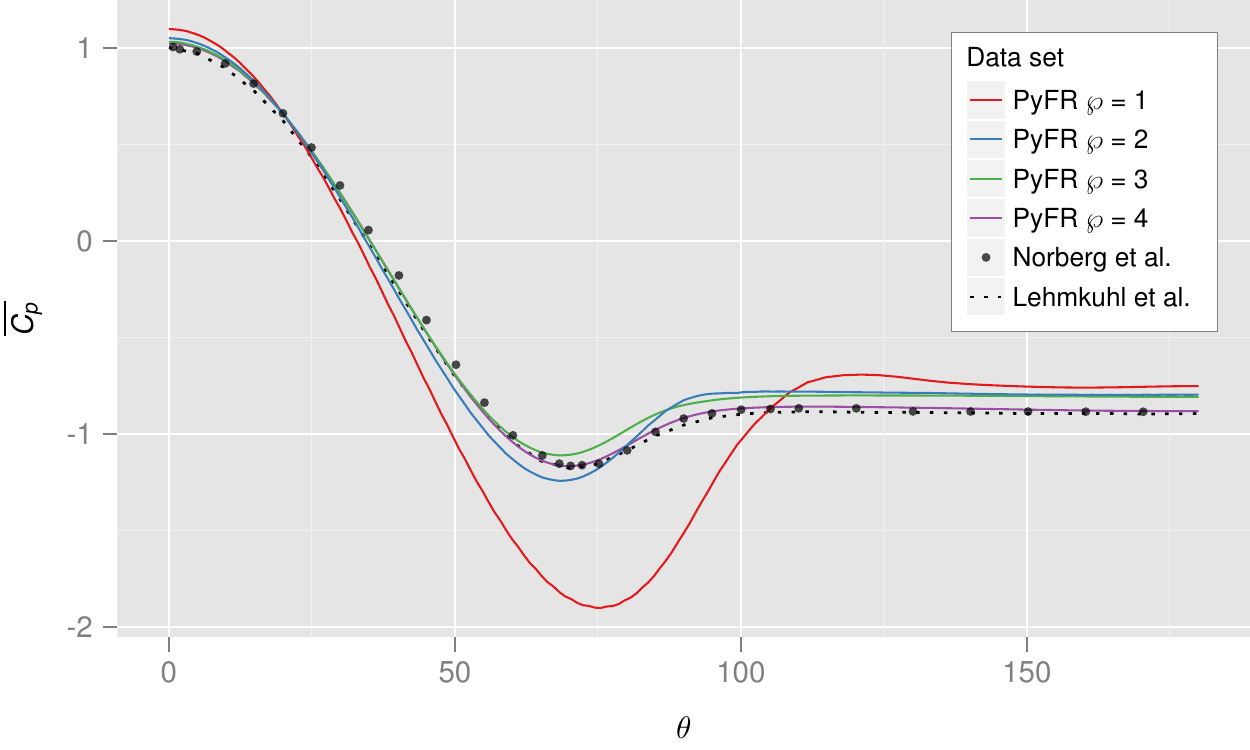}%
  \caption{\label{fig:cp}Averaged pressure coefficient for $\wp =
    1,2,3,4$ compared with the experimental results of Norberg et
    al. \cite{norberg1988ldv} (from Kravchenko and Moin
    \cite{kravchenko2000numerical}) and DNS results of Lehmkuhl et
    al. \cite{lehmkuhlthe2013}.}
\end{figure}

Plots of averaged pressure coefficient $\overline{C_p}$ on the surface
of the cylinder are shown in \autoref{fig:cp}. The results are shown
alongside the experimental results of Norberg et al.
\cite{norberg1988ldv} (extracted from Kravchenko and Moin
\cite{kravchenko2000numerical}), and the Mode-L DNS results of Lehmkuhl
et al. \cite{lehmkuhlthe2013}. At $\wp=1$ we observe a large negative
pressure coefficient near the top and bottom of the cylinder, which
includes the location of maximum skin friction coefficient
\cite{breuer1998large}. However, with increasing $\wp$ the results tend
towards those of the previous studies.

The averaged pressure coefficient at the base of the cylinder
$\overline{C_{pb}}$, and the averaged separation angle $\theta_s$
measured from the leading stagnation point are tabulated in
\autoref{tab:cylinder-results}, along with experimental data from
Norberg et al. \cite{norberg1988ldv} at a similar $Re = 4020$
(extracted from Kravchenko and Moin \cite{kravchenko2000numerical}),
experimental data from Parnaudeau et al.
\cite{parnaudeau1998experimental}, and Mode-L DNS data from Lehmkuhl et
al. \cite{lehmkuhlthe2013}. Once again, with increasing $\wp$ the
results tend towards those of the previous studies.

\begin{table}
  \centering
  \caption{\label{tab:cylinder-results}Comparison of quantitative
    values with experimental and DNS results.}
  \begin{tabular}{rrr} \toprule
    & \ccol{$-\overline{C_{pb}}$} &
    \ccol{$\theta_s /{}^\circ$} \\
    \midrule
    PyFR pri/tet $\wp = 1$ & $0.752$ & $92$ \\
    PyFR pri/tet $\wp = 2$ & $0.796$ & $89$ \\
    PyFR pri/tet $\wp = 3$ & $0.808$ & $88$ \\
    PyFR pri/tet $\wp = 4$ & $0.881$ & $88$ \\
    \midrule
    Parnaudeau et al. \cite{parnaudeau1998experimental} &  &
    $88$ \\
    Lehmkuhl et al. \cite{lehmkuhlthe2013} & $0.877$ & $87.8$
    \\
    Norberg et al. \cite{norberg1988ldv,kravchenko2000numerical} &
    $0.880$ & \\
    \bottomrule
  \end{tabular}
\end{table}

\begin{figure}
  \centering
  \includegraphics{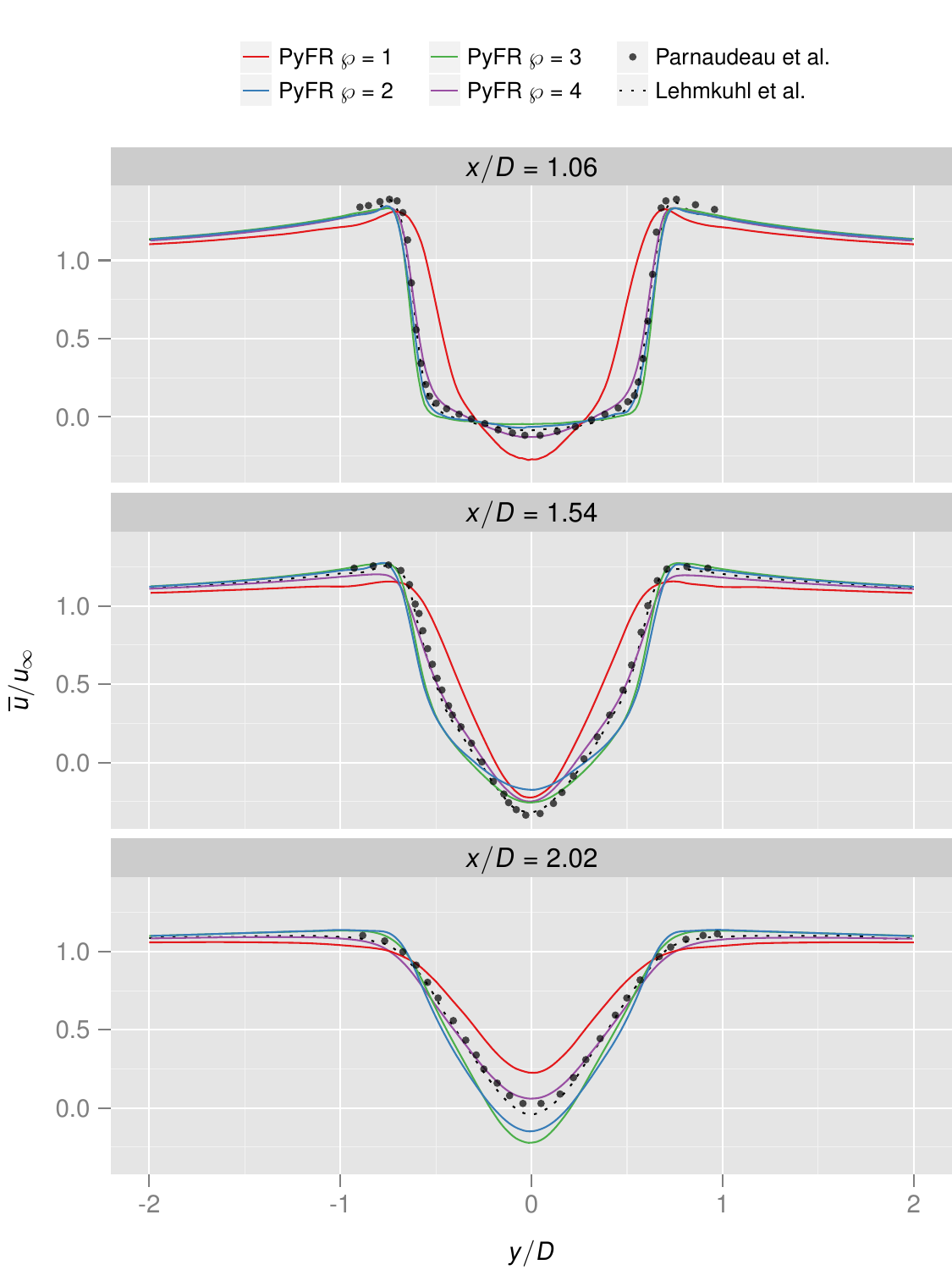}%
  \caption{\label{fig:uprofiles}Time-span-average stream-wise velocity
    for $\wp = 1,2,3,4$ compared with the experimental results of
    Parnaudeau et al. \cite{parnaudeau1998experimental} and DNS results
    of Lehmkuhl et al. \cite{lehmkuhlthe2013}.}
\end{figure}

Plots of averaged stream-wise velocity at $x/D = 1.06$, $1.54$, and
$2.02$ are shown in \autoref{fig:uprofiles}. The results are shown
alongside the experimental results of Parnaudeau et al.
\cite{parnaudeau1998experimental} and the Mode-L DNS results of
Lehmkuhl et al. \cite{lehmkuhlthe2013}. With $\wp=1$ the profiles
deviate significantly from the previous studies. On increasing the
order to $\wp=2$ the results are improved. We observe a U-shape profile
at $x/D = 1.06$, with strong gradients near the mixing layer between
the wake and the free stream. The $\wp=4$ results agree well those of
the previous studies. The only exception is the reduced magnitude of
the averaged velocity deficit near the centre of the wake at $x/D =
1.54$.

\begin{figure}
  \centering
  \includegraphics{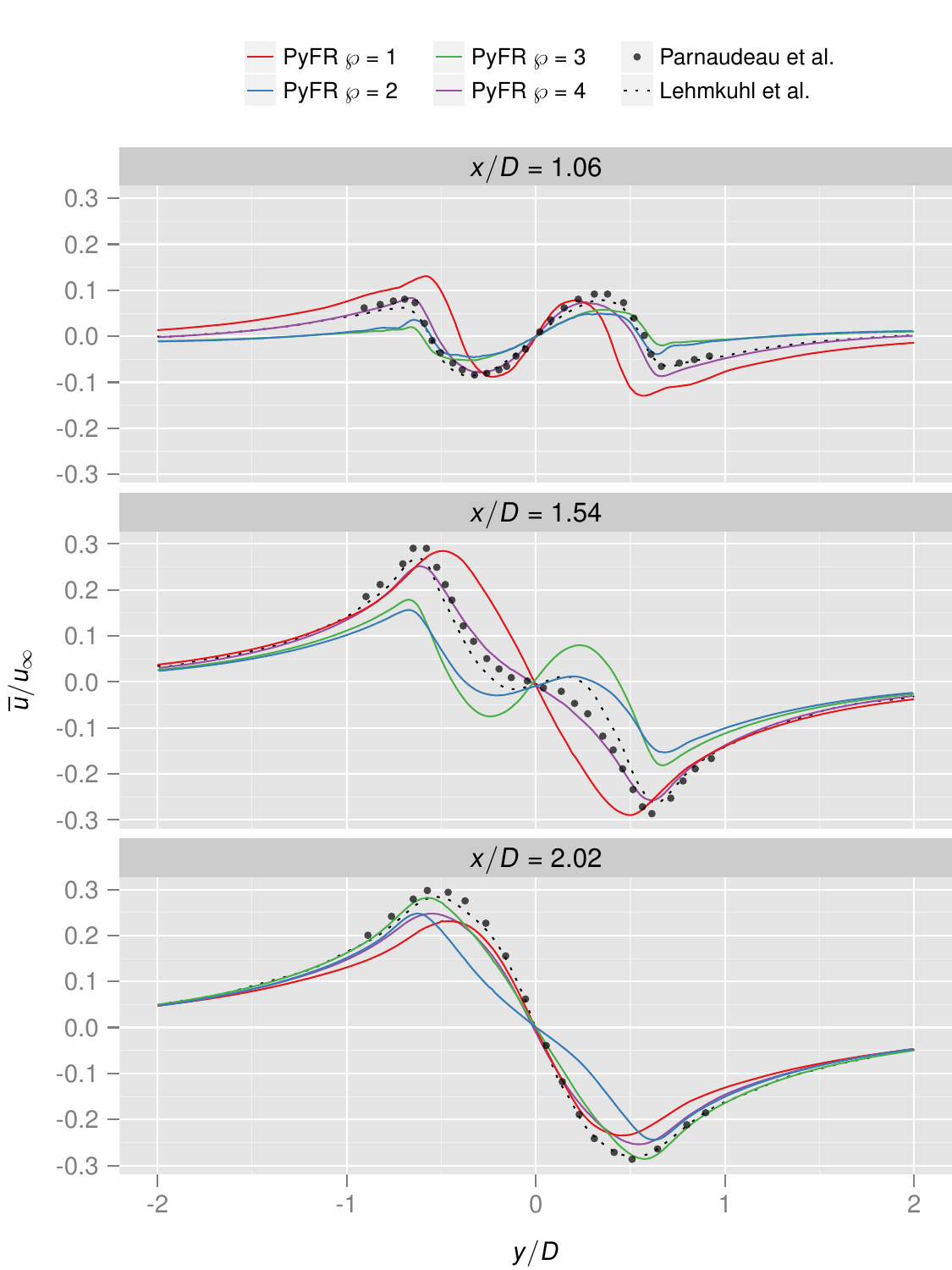}%
  \caption{\label{fig:vprofiles}Time-span-average cross-stream velocity
    for $\wp = 1,2,3,4$ compared with the experimental results of
    Parnaudeau et al. \cite{parnaudeau1998experimental} and DNS results
    of Lehmkuhl et al. \cite{lehmkuhlthe2013}.}
\end{figure}

Plots of averaged cross-wise velocity at $x/D = 1.06$, $1.54$, and
$2.02$ are shown in \autoref{fig:vprofiles}. The results are also shown
alongside the experimental results of Parnaudeau et al.
\cite{parnaudeau1998experimental} and the Mode-L DNS results of
Lehmkuhl et al. \cite{lehmkuhlthe2013}. The $\wp=4$ results agree well
with those of the previous studies.

\section{Conclusions}
\label{sec:conclusions}

In this paper we have demonstrated the ability of PyFR to perform
high-order accurate unsteady simulations of flow on mixed unstructured
grids using \emph{heterogeneous} multi-node hardware. Specifically,
after benchmarking single-node performance for various platforms, PyFR
v0.2.2 was used to undertake simulations of unsteady flow over a
circular cylinder at Reynolds number $3\,900$ using a mixed unstructured
grid of prismatic and tetrahedral elements on a desktop workstation
containing an Intel Xeon E5-2697 v2 CPU, an NVIDIA Tesla K40c GPU, and
an AMD FirePro W9100 GPU. Results demonstrate that PyFR achieves
performance portability across various hardware platforms. In
particular, the ability of PyFR to target individual platforms with
their `native' language leads to significantly enhanced performance
\emph{cf.} targeting each platform with OpenCL alone. PyFR was also
found to be performant on the heterogeneous multi-node system achieving
a significant fraction of the available FLOP/s.  Further, the numerical
results obtained using a mixed unstructured grid of prismatic and
tetrahedral elements were found to be in good agreement with previous
experimental and numerical data.

\section*{Acknowledgements}

The authors would like to thank the Engineering and Physical Sciences
Research Council for their support via a Doctoral Training Grant and
an Early Career Fellowship (EP/K027379/1). The authors would also like
to thank AMD, Intel, and NVIDIA for hardware donations.

\end{document}